\documentclass[cits]{PoS}
\usepackage{multirow, longtable, lscape}
\usepackage{amsmath,amsthm}
\usepackage[lofdepth,lotdepth]{subfig}
\usepackage{array}
\usepackage[font=small]{caption}

\graphicspath{{./figs/}}

\providecommand{\wbar}[1]{\overline#1}
\providecommand{\abs}[1]{\lvert#1\rvert}

\providecommand{\ket}[1]{\lvert#1\rangle}
\providecommand{\matrixe}[3]{\langle#1\lvert#2\rvert#3\rangle}
\providecommand{\expv}[1]{\langle#1\rangle}

\definecolor{1st}{RGB}{255,165,0}
\definecolor{2nd}{RGB}{135,206,250}
\definecolor{3rd}{RGB}{173,255,47}

\newcolumntype{C}[1]{>{\centering\let\newline\\\arraybackslash\hspace{0pt}}m{#1}}
\title{Current Status of Indirect CP Violation in Neutral Kaon System }

\ShortTitle{$\varepsilon_K$} 

\author{\speaker{Yong-Chull Jang}, and Weonjong Lee\\
        Lattice Gauge Theory Research Center, CTP, and FPRD,\\
        Department of Physics and Astronomy, \\
        Seoul National University, Seoul,       
        151-747, South Korea\\
        E-mail: \email{wlee@snu.ac.kr}}

\author{SWME Collaboration
}

\abstract{ In the standard model (SM), the CP violation is introduced
  through a single phase in the CKM matrix.
  The neutral kaon system is one of the most precise channels to test
  how the SM theory describes the experiment data such as $\epsilon_K$
  accurately.
  The indirect CP violation is parametrized into $\epsilon_{K}$, which
  can be calculated directly using lattice QCD.
  In this calculation, the largest uncertainty comes from two sources:
  one is $\hat{B}_K$ and the other is $V_{cb}$. 
  We use the lattice results of $\hat{B}_K$ and exclusive $V_{cb}$ to
  calculate the theoretical estimate of $\epsilon_K$, which turns
  out to be $3.1\sigma$ away from its experimental value.
  Here, the error is evaluated using the standard error propagation
  method.}

\FullConference{The 30th International Symposium on Lattice Field Theory\\
                 June 24 -- 29,  2012\\
                 Cairns, Australia}

\begin{document}

\section{Introduction}
  The neutral kaon system has two kinds of CP violation: indirect and direct
  CP violation. 
  Indirect CP violation is parametrized into $\epsilon_{K}$.
%
%  In the standard model (SM), it is the phase factor in the CKM
%  matrix~\cite{Cabibbo1963:PhysRevLett.10.531,
%    Kobayashi1973:ProgTheorPhys.49.652} that causes the CP
%  violation.
%
%  The CKM matrix has one phase $\delta \ne 0$ which produces CP
%  violation\footnote{In the Wolfenstein parametrization, the imaginary 
%    part $\bar{\eta} \neq 0$ is responsible for the CP violation.}.
%
%  We will provide the relationship between $\epsilon_{K}$ and $\delta$
%  (the phase factor of CKM matrix) or $\bar{\eta}$ in this paper later.
%
  The experimental value of $\epsilon_{K}$ is very well known 
  \cite{Nakamura2010:RevPtlPhys}:
  \begin{align}
  \epsilon_{K} & = (2.228 \pm 0.011) \times 10^{-3} 
  \times e^{i\phi_\epsilon}\,, \qquad 
  \phi_\epsilon = 43.51 \pm 0.05 .
  \end{align}
  We can also calculate $\epsilon_{K}$ directly from the SM using
  $B_K$, $V_{cb}$, and other input parameters, which are determined from
  other experiments and the SM theory.
  By comparing these two values, we can test a fundamental ansatz of
  the SM, the unitarity of the CKM matrix.

  Here, we calculate $\epsilon_{K}$ directly from the SM using the
  known parameters with their errors in control.
  Two of the most important input parameters are $\hat{B}_{K}$ and
  $V_{cb}$, which dominate the statistical and systematic uncertainty
  in $\epsilon_K$.
  During a past decade, lattice QCD has reduced the $B_K$ error
  dramatically down to less than 5\% level
  \cite{swme-prl-1,rbc-prd-2011-1,alv-prd-2010-1} as well as $V_{cb}$
  \cite{fnal/milc-pos-2010-1}.

  There are two independent methods to determine $V_{cb}$: inclusive 
  channels and exclusive channel.
  There exists about $2~3\sigma$ difference in $V_{cb}$ between
  inclusive and exclusive methods.
  We address this issue on how this have an effect on $\epsilon_K$.

%  The CKM matrix $V_{ij}$ is an unitary matrix.
  % 
%  Hence, in a popular parametrization method known as Wolfenstein
%  parameters \cite{Buras2008:PhysRevD.78.033005}, this unitarity
%  ansatz is used to constrain the parameters.
  %
  Here, we use the Wolfenstein parameters for the CKM matrix to
  calculate $\epsilon_K$ mainly because they are convenient.

  Let us define a parameter $\Delta$ which test the unitarity ansatz
  directly as
  \[
  \Delta \equiv V \cdot V^\dagger - I .
  \]
  Here, we calculate $\epsilon_K$ and $\Delta$ to test the SM.
  The CKM matrix elements are accurate up to $\mathcal{O}(\lambda^{5})
  \approx 3 \times 10^{-4}$.
  We use the standard error propagation method to estimate the errors.

\section{Review of the Neutral Kaon Mixing: $\epsilon_{K}$}
\label{sec:KaonMixingReview}
  The neutral kaon system forms a two dimensional Hilbert space.
  In this subspace, the time evolution of the neutral kaon state vectors
  can be described by the effective Hamiltonian $\hat{H}_{\text{eff}}$
\begin{equation}
  i\frac{d}{dt} \ket{K(t)} = \hat{H}_{\text{eff}} \ket{K(t)}
  ,\quad 
  \hat{H}_{\text{eff}} = \hat{M} - i\frac{\hat{\Gamma}}{2} .
\end{equation} 
  $\hat{M}$ $(\hat{\Gamma})$ is the dispersive (absorptive) part.
  Here, the dispersive part gives the mass eigenvalues of $K_S$ and
  $K_L$, and the absorptive part represents the decay rates of $K_S$
  and $K_L$.
  Let us take the basis with the CP even $\ket{K_{1}}$ and odd
  $\ket{K_{2}}$ states,
\begin{gather}
  \ket{K_{1}} = \frac{1}{\sqrt{2}} 
  \big( \ket{K^{0}} + \ket{\wbar{K^{0}}} \big) 
  ,\quad
  \ket{K_{2}} = \frac{1}{\sqrt{2}} 
  \big( \ket{K^{0}} - \ket{\wbar{K^{0}}} \big) .
\end{gather}

In this basis, the matrix elements can be written as the followings:
\begin{equation}
  M =
  \begin{pmatrix} 
    M_{1} & +im^{\prime}\\
    -im^{\prime} & M_{2}
  \end{pmatrix}
  ,\;
  \Gamma =
  \begin{pmatrix} 
    \Gamma_{1} & +i\gamma^{\prime} \\
    -i\gamma^{\prime} & \Gamma_{2}
  \end{pmatrix} .
\end{equation}
By construction of the formal perturbation theory in quantum field
theory known as the Wigner-Weisskopf theory
\cite{Christ2011:hep-lat/1201.2065}, $M$ and $\Gamma$ are hermitian
matrices.
In addition, if we assume that the CPT invariance is exactly
respected, then $m^{\prime}$ and $\gamma^{\prime}$ must be real.

Solving the eigenvalue problem with this matrix representation for
$H_{\text{eff}}$, the eigenstates
\begin{gather}
  \ket{K_{S}} = \frac{1}{ \sqrt{ 1+\abs{\tilde{\epsilon}}^{2} } } 
  (\ket{K_{1}} + \tilde{\epsilon} \ket{K_{2}})
  ,\quad
  \ket{K_{L}} = \frac{1}{ \sqrt{ 1+\abs{\tilde{\epsilon}}^{2} } } 
  (\ket{K_{2}} + \tilde{\epsilon} \ket{K_{1}})
\end{gather}
have the mass $M_{S,L}$ and the decay rate $\Gamma_{S,L}$, respectively.
The small CP impurity $\tilde{\epsilon}$ satisfies the following equation
\begin{equation}
  \tilde{\epsilon} = \tilde{\epsilon}_{(0)} 
  ( 1 + \tilde{\epsilon}^{2} ) .
\end{equation}
Here, the $\tilde{\epsilon}_{(0)}$ parameter is defined as
\begin{equation}
  \label{eq:epsilon tilde 0}
  \tilde{\epsilon}_{(0)} 
  \equiv \frac{ -i \big( m^{\prime}-\frac{i}{2}\gamma^{\prime} \big) }{
    \big( M_{1} - M_{2} \big) -\frac{i}{2} \big( \Gamma_{1} - \Gamma_{2}
    \big) }
  = e^{i\theta} \sin{\theta} ( \frac{m^{\prime}}{\Delta M_{K}} -
  i\cot{\theta} \frac{\gamma^{\prime}}{\Delta\Gamma_{K}} )
  + \mathcal{O}\big( {\tilde{\epsilon}_{(0)}}^{3} \big) ,
\end{equation}
where
\begin{equation}
  \Delta M_{K} = M_{L} - M_{S} 
  ,\;
  \Delta\Gamma_{K} = \Gamma_{S}-\Gamma_{L} 
  ,\;
  \tan{\theta} = \frac{2\Delta M_{K}}{\Delta\Gamma_{K}} .
\end{equation}
The solution of $\tilde{\epsilon}$ can be obtained by iteration.
Since $\tilde{\epsilon}_{(0)}$ is of the order of $10^{-3}$,
we may expand $\tilde{\epsilon}$ perturbatively as follows, 
\begin{equation}
  \tilde{\epsilon}
  = \tilde{\epsilon}_{(0)} + {\tilde{\epsilon}_{(0)}}^{3} +
  2{\tilde{\epsilon}_{(0)}}^{5} + 5{\tilde{\epsilon}_{(0)}}^{7} + \cdots .
\end{equation}

In Eq.~\eqref{eq:epsilon tilde 0}, $M_{1,2}$ and $\Gamma_{1,2}$ can be
safely replaced by the eigenvalues $M_{S,L}$ and $\Gamma_{S,L}$ which 
are experimental observables.
Note that this approximation makes an error of the size
$\mathcal{O}({\tilde{\epsilon}_{(0)}}^{3}) \approx 10^{-9}$ which is
of no interest to us.
In the case of $\gamma^{\prime}/\Delta\Gamma_{K}$, we presume the
following assumptions:
\begin{itemize}
\item First, we make the approximation
  $\Delta\Gamma_{K} \cong \Gamma_{1}$ 
  which is good up to the precision of $10^{-3}$.
\item Second, we assume that the contribution from the two pion state
  is dominant in $\gamma^{\prime}, \Gamma_{1}$ which is good in the
  precision level of $10^{-3}$.
\item Third, we assume that the contribution from the $I=0$ two pion
  state is dominant in $\gamma^{\prime}, \Gamma_{1}$ compared with
  that from the $I=2$ state. This approximation is good up to the
  precision of $10^{-7}$.
\end{itemize}
Using these assumptions, we can approximate the 
$\gamma^{\prime}/\Delta\Gamma_{K}$ as follows,
%
%\begin{equation}\label{eq:xi0}
%  \frac{\gamma^{\prime}}{\Delta\Gamma_{K}} \cong
%  \frac{\mathrm{Im}A_{0}}{\mathrm{Re}A_{0}} \equiv \xi_{0}.
%\end{equation}
\begin{equation}\label{eq:xi0}
  \frac{\gamma^{\prime}}{\Delta\Gamma_{K}} 
  = \xi_{0} + \mathcal{O}\big(10^{-7}\big),\quad
  \xi_{0} \equiv \frac{\mathrm{Im}A_{0}}{\mathrm{Re}A_{0}} .
\end{equation}
Then, we can express $\epsilon_{K}$ approximately as follows,
\begin{align}\label{eq:epsilon_K}
  \epsilon_{K} & \equiv
  \frac{ \matrixe{ \pi\pi(I=0) }{ H_{W} }{ K_{L} } }{ \matrixe{ \pi\pi(I=0)
    }{ H_{W} }{ K_{S} } }
  = \tilde{\epsilon}_{(0)} + i \xi_{0} 
  + \mathcal{O}\big({\tilde{\epsilon}_{(0)}}^{3} \big) 
  = e^{i\theta} \sin{\theta} \Big( \frac{m^{\prime}_{(6)}}{\Delta M_{K}} +
       \xi_{0} \Big) + \Delta \epsilon_K\,. 
%       \\
%  \Delta \epsilon_K & \equiv \epsilon_{K} - 
%  e^{i\theta} \sin{\theta} \Big( \frac{m^{\prime}}{\Delta M_{K}} +    
%       \xi_{0} \Big) 
\end{align}

The correction of $\mathcal{O}( {\tilde{\epsilon}_{(0)}}^{3} )$
in Eq.~\eqref{eq:epsilon_K} is smaller than both the current
experiment precision and the size of the long distance contributions
of the $m^{\prime}$ \cite{Buras2010:PhysLettB.688}.
%
%Hence, we neglect this term and its higher order terms.
%
The last expression in Eq.~\eqref{eq:epsilon_K} is obtained by
substituting $\tilde{\epsilon}_{(0)}$ with Eq.~\eqref{eq:epsilon tilde 0}
and Eq.~\eqref{eq:xi0}.
The correction of $\Delta \epsilon_K$ contains both short-distance
(SD) contribution and long-distance (LD) contribution, which are
expected to be about $\approx 5\%$\cite{Buras2010:PhysLettB.688}.
Here, we also neglect this contribution from $\Delta \epsilon_K$,
mainly because it is not known to a sufficient precision
theoretically.

In this analysis, we take into account only the short-distance
contribution from the dimension 6 operators $m^{\prime}_{(6)}$.
In the SM, this part can be calculated from the box
diagram~\cite{Buras1998:hep-ph/9806471}.
Here, we follow the notations in~\cite{Buras1998:hep-ph/9806471}.
Then, we can obtain the following master formula which will be used in
this analysis:
\begin{equation}\label{eq:epsK^SM}
  \abs{\epsilon_{K}^{\text{SM}}} 
  = \sqrt{2}\sin{\theta} \Big( C_{\epsilon} \hat{B}_{K} X 
  + \xi_{0} \Big) ,
\end{equation}
where
\begin{align}
  X &= \bar{\eta}\lambda^{2} \abs{V_{cb}}^{2} \times 
  \Big[ \abs{V_{cb}}^{2} (1-\bar{\rho})
    \eta_{2} S_{0}(x_{t}) 
    + \eta_{3} S_{0}(x_{c},x_{t}) - \eta_{1} S_{0}(x_{c}) \Big] \\
  C_{\epsilon} &= \frac{ G_{F}^{2} F_K^{2} m_{K^{0}} M_{W}^{2} }{ 6\sqrt{2}
    \pi^{2} \Delta M_{K} } .
\end{align}
where we use the experimental value for $\Delta M_{K}$.
The input $\xi_{0}$ has been taken from the lattice calculation which
accounts the long distance contribution.
\begin{table}[h!]\scriptsize
\centering
\renewcommand{\arraystretch}{1.1}
\subfloat[][]{
%\resizebox{7cm}{!}{
\begin{tabular}{C{1cm}|p{3.5cm}|p{0.5cm}}
\hline\hline
%\multicolumn{1}{c}{Parameter}&
%\multicolumn{1}{|c|}{Value}&
%\multicolumn{1}{c}{Reference}\\ [0.2ex]
%\hline\hline
$G_{F}$ & $1.16637(1) \times 10^{-5}$ GeV$^{-2}$
&\cite{Nakamura2010:RevPtlPhys} \\ \cline{1-3}
$M_{W}$ & 80.399(23) GeV 
&\cite{Nakamura2010:RevPtlPhys} \\ \cline{1-3}
$m_{c}(m_{c})$ & $1.25(9)$ GeV 
&\cite{Buras2008:PhysRevD.78.033005} \\ \cline{1-3}
$m_{t}(m_{t})$ & $162.7(1.3)$ GeV 
&\cite{Buras2008:PhysRevD.78.033005} \\ \cline{1-3}
$\eta_{1}$ & $1.43(23)$ 
&\cite{Buras2008:PhysRevD.78.033005} \\ \cline{1-3}
$\eta_{2}$ & $0.5765(65)$ 
&\cite{Buras2008:PhysRevD.78.033005} \\ \cline{1-3}
$\eta_{3}$ & $0.47(4)$ 
&\cite{Buras2008:PhysRevD.78.033005} \\ \cline{1-3}
$\theta$ & $43.51(5)^{\circ}$
&\cite{Nakamura2010:RevPtlPhys} \\ \cline{1-3}
$m_{K^{0}}$ & $497.614(24)$ MeV 
&\cite{Nakamura2010:RevPtlPhys} \\ \cline{1-3}
$\Delta M_{K}$ &  $3.483(6) \times 10^{-12}$ MeV &\cite{Nakamura2010:RevPtlPhys} \\
\hline\hline
\end{tabular} 
%}
\label{table:SM param}
}\qquad
\subfloat[Wolfenstein Parameters][Wolfenstein Parameters]{
%\resizebox{.45\textwidth}{!}{
\begin{tabular}{C{0.7cm}|p{1.5cm}|p{2cm}}
\multicolumn{1}{c}{}&\multicolumn{1}{c}{}&\multicolumn{1}{c}{} \\
\multicolumn{1}{c}{}&\multicolumn{1}{c}{}&\multicolumn{1}{c}{} \\
\hline\hline
%\multicolumn{1}{c}{Parameter}&
%\multicolumn{1}{|c|}{Value}&
%\multicolumn{1}{c}{Reference}\\ [0.2ex]
%\hline\hline
\multirow{2}{*}{$A$} & $0.808(22)$ 
&\cite{Nakamura2010:RevPtlPhys} CKMfitter \\ \cline{2-3}
& $ 0.832(17)$ 
&\cite{Nakamura2010:RevPtlPhys} UTfit \\ \cline{1-3}
\multirow{2}{*}{$\lambda$} & $0.2253(7)$ 
&\cite{Nakamura2010:RevPtlPhys} CKMfitter \\ \cline{2-3}
& $ 0.2246(11)$ 
&\cite{Nakamura2010:RevPtlPhys} UTfit \\ \cline{1-3}
\multirow{2}{*}{$\bar{\rho}$} & $\displaystyle 0.132^{+0.022}_{-0.014}$ &\cite{Nakamura2010:RevPtlPhys} CKMfitter \\ \cline{2-3}
& $0.130(18)$ 
&\cite{Nakamura2010:RevPtlPhys} UTfit \\ \cline{1-3}
\multirow{2}{*}{$\bar{\eta}$} & $0.341(13)$ 
&\cite{Nakamura2010:RevPtlPhys} CKMfitter \\ \cline{2-3}
& $0.350(13)$ 
&\cite{Nakamura2010:RevPtlPhys} UTfit \\
\hline\hline
\end{tabular}
%}
\label{table:Wolf param}
}\\
\subfloat[Lattice Calculation][Lattice Calculation]{
%\resizebox{.4\columnwidth}{!}{
\begin{tabular}{C{0.6cm}|p{3.1cm}|p{1.8cm}}
\multicolumn{1}{c}{}&\multicolumn{1}{c}{}&\multicolumn{1}{c}{} \\
\hline\hline
%\multicolumn{1}{c}{Parameter}&
%\multicolumn{1}{|c|}{Value}&
%\multicolumn{1}{c}{Reference}\\ [0.2ex]
%\hline\hline
\multirow{2}{*}{$F_K$} & $156.1(0.2)(0.8)(0.2)$ MeV
&\cite{Nakamura2010:RevPtlPhys}\\ \cline{2-3}
& $156.1(1.1)$ MeV 
&\cite{Laiho2009:PhysRevD.81.034503} LAT.AVG.\\ \cline{1-3}
\multirow{2}{*}{$\hat{B}_{K}$} & $0.7674(99)$ &\cite{Laiho2009:PhysRevD.81.034503} LAT.AVG.\\ \cline{2-3}
& $0.727(4)(38)$ 
&\cite{swme-prl-1} SWME\\ \cline{1-3}
$\xi_{0}$ & $-1.63(19)(20) \times 10^{-4}$ &\cite{Blum2011:PhysRevLett.108.141601}\\
\hline\hline
\end{tabular}
%}
\label{table:Lat param}
}\qquad
\subfloat[$V_{cb}$][$V_{cb}$]{
%\resizebox{4cm}{!}{
\begin{tabular}{C{0.7cm}|p{3cm}|p{2.3cm}}
\hline\hline
%\multicolumn{1}{c}{Parameter}&
%\multicolumn{1}{|c|}{Value}&
%\multicolumn{1}{c}{Reference}\\ [0.2ex]
%\hline\hline
\multirow{6}{*}{$\abs{V_{cb}}$} & $41.85(42)(9)(59) \times 10^{-3}$ &\cite{Laiho2011:hep-ex/1107.3934} $(X_{c}l\nu + X_{s}\gamma)_{\text{Kin}}$ \\ \cline{2-3}
& $41.68(44)(9)(58) \times 10^{-3}$ 
&\cite{Laiho2011:hep-ex/1107.3934} $(X_{c}l\nu)_{\text{Kin}}$ \\ \cline{2-3}
& $41.87(25) \times 10^{-3}$ 
&\cite{Laiho2011:hep-ex/1107.3934} $(X_{c}l\nu + X_{s}\gamma)_{\text{1S}}$ \\ \cline{2-3}
& $42.31(36) \times 10^{-3}$ 
&\cite{Laiho2011:hep-ex/1107.3934} $(X_{c}l\nu)_{\text{1S}}$ \\ \cline{2-3}
& $41.5(7) \times 10^{-3}$ 
&\cite{Nakamura2010:RevPtlPhys} Incl.PDG.AVG.\\ \cline{2-3}
& $39.5(1.0) \times 10^{-3}$ 
&\cite{Laiho2009:PhysRevD.81.034503} Excl. \\
\hline\hline
\end{tabular}
%}
\label{table:Vcb}
}
\caption{Input Parameters}
\label{table:Input Parameters}
\end{table}
\section{Input Parameters}
The parameters, $m_{c}, m_{t}, \eta_{1}, \eta_{2}, \eta_{3}$ depend on
the renormalization scheme, and so are taken from the single reference
for consistency (Table~\ref{table:SM param}\footnote{ In
  Ref.~\cite{Brod2011:PhysRevLett.108.121801}, they reported results
  of $\eta_1$ up to NNLO but end up with a noticeably larger error
  bar. Hence, we decide to use the NLO value.}).
The CKMfitter and UTfit results in Table~\ref{table:Wolf param} are
obtained by their own global fit method using the same PDG inputs.

In $\hat{B}_{K}$ calculation in Table~\ref{table:Lat param}, BMW quotes
the smallest systematic error.
It dominates the smallness of the lattice average error. RBC-UKQCD
collaboration calculates $\mathrm{Im}A_{2}$ on the lattice. 
Using this value, they determine $\xi_{0}$ through the relation
\begin{equation}
  \mathrm{Re} \Big( \frac{\epsilon^\prime_{K}}{\epsilon_{K}} \Big) =
  \frac{1}{\sqrt{2}\abs{\epsilon_{K}}}
  \frac{\mathrm{Re}A_{2}}{\mathrm{Re}A_{0}}
  \Big( \frac{\mathrm{Im}A_{2}}{\mathrm{Re}A_{2}} - \xi_{0} \Big) .
\end{equation}
Other inputs such as $\mathrm{Re}A_{0}, \mathrm{Re}A_{2},
\epsilon_{K},$ and ${\epsilon^{\prime}_{K}}/{\epsilon_{K}}$ are taken
from experiments.

Inclusive $V_{cb}$ can be extracted from global fit of measured
moments (lepton energy, hadronic mass, and photon energy) of the decay
channels:
\begin{equation*}
  B \rightarrow X_{c}l\nu,\; \qquad B \rightarrow X_{s}\gamma .
\end{equation*}
%
% Theoretical fitting functions are obtained by applying OPE to the decay
% amplitude in $\alpha_{s}$ and $\Lambda/m_{b}$.
%
% There are two $\Lambda/m_{b}$ expansion schemes, 1S and kinetic,
% depending on the defining method for the threshold mass.
%
% For each choice of analysis channels, we tabulate the value separately
% through the first four rows of [Table~\ref{table:Vcb}].
%
% Also quotes the PDG average value for the inclusive results.
%
We use the PDG average value as the representative of the inclusive
$V_{cb}$.
The quoted exclusive $V_{cb}$ is the average of two semi-leptonic
decay channels:
\begin{equation*}
  B \rightarrow D^*\ell\nu\,, \qquad B \rightarrow D\ell\nu \,.
\end{equation*}
For each scalar and vector channel, HFAG result is combined with
FNAL/MILC lattice QCD calculation of the zero recoil form factor.

\section{Error Estimate}
For the function with $N$ arguments, the error propagation formula gives
the combined error $\sigma_{f}$ in terms of the errors of each arguments:
\begin{equation}
  \sigma_{f}^{2} = \sum_{j,k=1}^{N} C_{jk} \frac{\partial
    f(\mathbf{x})}{\partial x_{j}} \bigg\vert_{\expv{\mathbf{x}}}
  \frac{\partial f(\mathbf{x})}{\partial x_{k}}
  \bigg\vert_{\expv{\mathbf{x}}} \sigma_{x_{j}} \sigma_{x_{k}} ,
\end{equation}
where $C_{jk}$ denotes the normalized correlation between the
parameters $x_{j}$ and $x_{k}$, and $|C_{ij}| \leq 1$.
Especially the diagonal components $C_{ii} = 1$.
We turn off the correlation and so $C_{ij} = \delta_{ij}$. 
In the case of asymmetric error, $\bar{\rho}$ given by CKMfitter, we
take a larger error and treat it as a symmetric error.

For $\epsilon_{K}^{\text{SM}}$,
\begin{align}
  f(\mathbf{x}) &= \abs{\epsilon_{K}(\mathbf{x})} \\
  \mathbf{x} &= (\theta, G_{F}, F_{K}, m_{K^{0}}, M_{W}, \Delta M_{K},
  \xi_{0}, \hat{B}_{K}, \lambda, \bar{\rho}, \bar{\eta}, \abs{V_{cb}},
  \eta_{1}, \eta_{2}, \eta_{3}, x_{c}, x_{t}) .
\end{align}
To check the unitarity of the CKM matrix,
\begin{align}
  f_{ij}(\mathbf{x}) &= [V(\mathbf{x})V^{\dag}(\mathbf{x}) - I]_{ij} 
  \equiv \Delta_{ij}  \\
  \mathbf{x} &= (A, \lambda, \rho, \eta; \abs{V_{cb}}) .
\end{align}
We use the Wolfenstein parametrization to evaluate each elements of the
CKM matrix $V_{ij}$, except for $V_{cb}$ itself. Real and imaginary part
of $\Delta_{ij}$ are separately treated.

%\newpage
\section{Results}
\begin{figure}[t]
\centering
\subfloat[CKMfitter][CKMfitter]{
\includegraphics[width=.31\columnwidth]{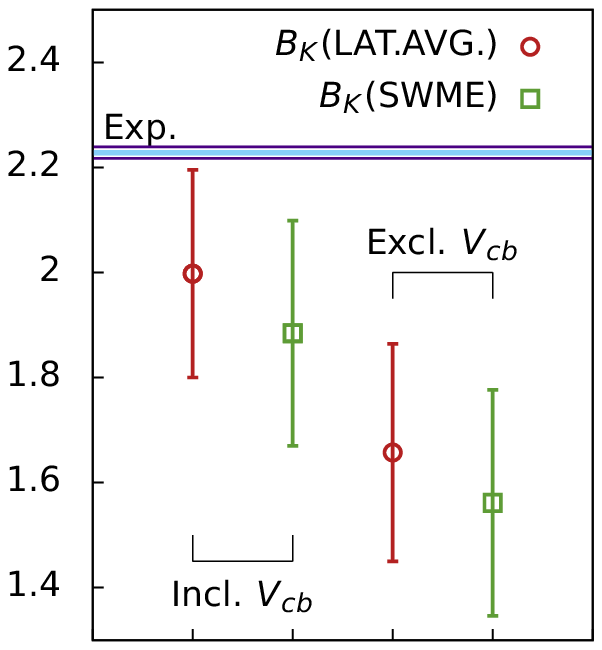}
\label{fig:epsK_CKMfitter}
}
\hspace{.05\columnwidth}
\subfloat[UTfit][UTfit]{
\includegraphics[width=.31\columnwidth]{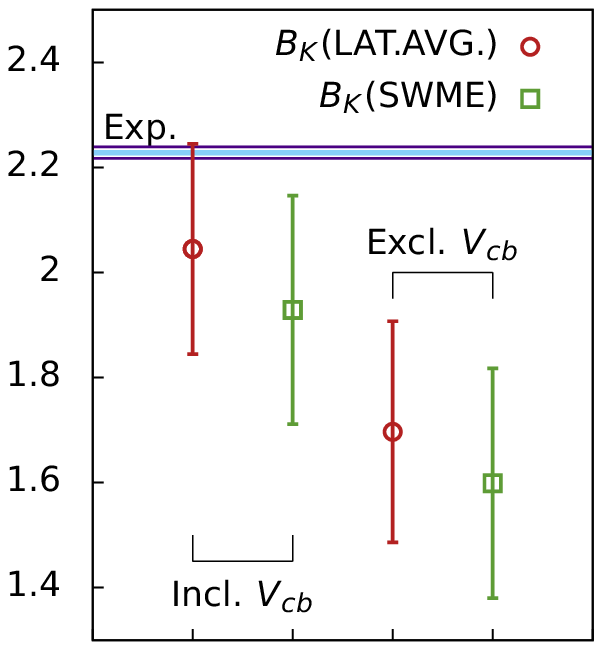}
\label{fig:epsK_UTfit}
}
\caption{$\epsilon_{K}(\times 10^{3})$}
\label{fig:epsilon_K}
\end{figure}

\begin{figure}[t]
\centering
\subfloat[$\mathrm{Re}\Delta_{22}$(CKMfitter)][$\mathrm{Re}\Delta_{22}$(CKMfitter)]{
\includegraphics[width=.21\columnwidth]{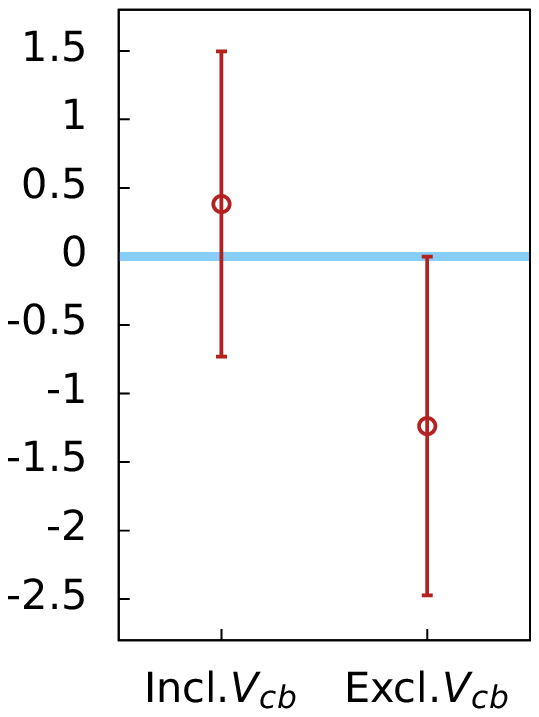}
\label{fig:Re_Delta22_CKMfitter}
}
\hspace{.02\columnwidth}
\subfloat[$\mathrm{Re}\Delta_{22}$(UTfit)][$\mathrm{Re}\Delta_{22}$(UTfit)]{
\includegraphics[width=.21\columnwidth]{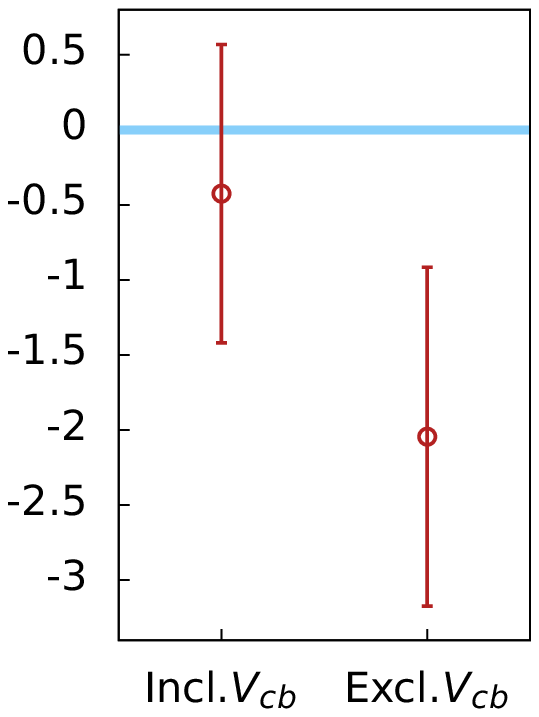}
\label{fig:Re_Delta22_UTfit}
}
\hspace{.02\columnwidth}
\subfloat[$\mathrm{Re}\Delta_{23}$(CKMfitter)][$\mathrm{Re}\Delta_{23}$(CKMfitter)]{
\includegraphics[width=.21\columnwidth]{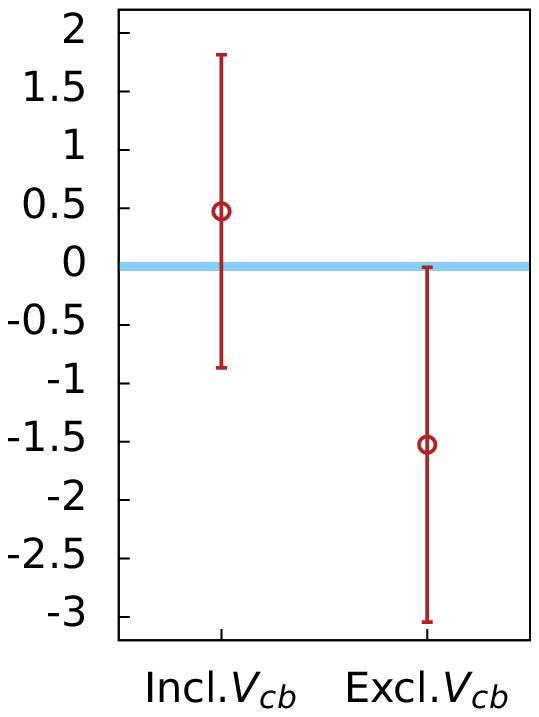}
\label{fig:Re_Delta23_CKMfitter}
}
\hspace{.02\columnwidth}
\subfloat[$\mathrm{Re}\Delta_{23}$(UTfit)][$\mathrm{Re}\Delta_{23}$(UTfit)]{
\includegraphics[width=.21\columnwidth]{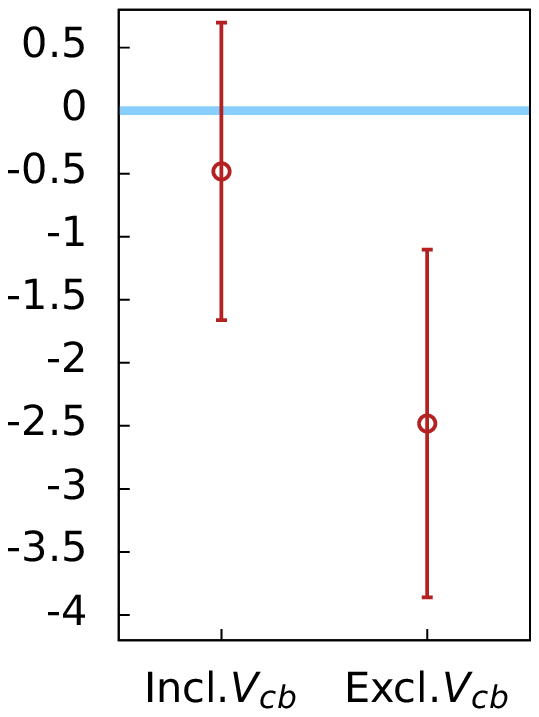}
\label{fig:Re_Delta23_UTfit}
}
\caption{$\mathrm{Re}\Delta_{22}(\times 10^{4})$ and $\mathrm{Re}\Delta_{23}(\times 10^{3})$ to test
  the CKM unitarity.}
\end{figure}

The Wolfenstein parameter set, $\hat{B}_{K}$, and $V_{cb}$ has a
multiple choice.
It forms 8 input parameter sets.
We calculate $\epsilon_{K}^{\text{SM}}$ for these sets as shown in
Fig.~\ref{fig:epsilon_K}.
We find out that $\abs{\epsilon_{K}}$ shows $3.1\sigma$ tension
between $\abs{\epsilon_{K}^{\text{Exp}}}$ and
$\abs{\epsilon_{K}^{\text{SM}}}$, using exclusive $V_{cb}$ and SWME
calculation of $B_K$ as shown in Fig.~\ref{fig:epsK_CKMfitter}.
With the UTfit Wolfenstein parameters, the tension is slightly reduced
to $2.9\sigma$ [Fig.~\ref{fig:epsK_UTfit}].
The deviation matrix $\Delta$ provides a test of the Wolfenstein
parameters, outputs of the global fit.
These are inputs for the $\epsilon_{K}^{\text{SM}}$, as well. So testing
the compatibility between the global fit results and the $V_{cb}$ is
needed to interpret the difference between $\epsilon_{K}^{\text{SM}}$ and
$\epsilon_{K}^{\text{SM}}$.
We find that the numerical size of $\Delta_{ij}$ has the following
hierarchy
\begin{equation}
  10^{2}\mathcal{O}\big(\abs{\mathrm{Re}\Delta_{12}}\big) =
  10\mathcal{O}\big(\abs{\mathrm{Re}\Delta_{22}}\big) =
  \mathcal{O}\big(\abs{\mathrm{Re}\Delta_{23}}\big)\,.
\end{equation}
$\mathrm{Re}\Delta_{22}$ shows that the difference between exclusive
$V_{cb}$ and Wolfenstein parameters from CKMfitter (UTfit) are about
$1.0\sigma$ ($1.8\sigma$) as shown in
Fig.~\ref{fig:Re_Delta22_CKMfitter} and \ref{fig:Re_Delta22_UTfit}.
$\mathrm{Re}\Delta_{23}$ shows the difference about
$1.0\sigma$ ($1.8\sigma$) as shown in 
Fig.~\ref{fig:Re_Delta23_CKMfitter} and \ref{fig:Re_Delta23_UTfit}.
Other components of $\Delta$, which does not depend on the choice of
inclusive or exclusive $V_{cb}$, are so small as to be consistent with
the unitary ansatz.
In Table~\ref{table:error fractions}, $V_{cb}$ dominates the error of
$\epsilon_{K}^{\text{SM}}$ regardless of the inclusive or exclusive
determination.
In case of the SWME calculation of $B_{K}$ and inclusive $V_{cb}$, both
contribute to the total error in comparable size.
In the case of the lattice average of $B_{K}$, $V_{cb}$ becomes an
extremely dominant error and the subdominant error comes from
$\bar\eta$.
\begin{table}[h]
\scriptsize 
\tabcolsep 3pt
\centering
\renewcommand{\arraystretch}{1.1}
\resizebox{.8\columnwidth}{!}{
\begin{tabular}{l|l|l|r|r|r|r|r|r|r|r|r}
\hline\hline
\multicolumn{1}{c}{W.P.}&
\multicolumn{1}{|c}{$V_{cb}$}&
\multicolumn{1}{|c}{$B_{K}$}&
\multicolumn{1}{|c}{$m_{c}$}&
\multicolumn{1}{|c}{$\eta_{1}$}&
\multicolumn{1}{|c}{$\eta_{3}$}&
\multicolumn{1}{|c}{$F_{K}$}&
\multicolumn{1}{|c}{$B_{K}$}&
\multicolumn{1}{|c}{$\xi_{0}$}&
\multicolumn{1}{|c}{$\bar{\rho}$}&
\multicolumn{1}{|c}{$\bar{\eta}$}&
\multicolumn{1}{|c}{$V_{cb}$} \\[0.2ex]
\hline\hline
\multirow{4}{*}{CKMfitter} & \multirow{2}{*}{Incl.} & LAT.AVG. & $10.62$ & $5.24$ & {\setlength{\fboxsep}{1.5pt}\colorbox{3rd}{$12.50$}} & $2.36$ & $1.98$ & $1.85$ & $4.29$ & {\setlength{\fboxsep}{1.5pt}\colorbox{2nd}{$17.29$}} & {\setlength{\fboxsep}{1.5pt}\colorbox{1st}{$41.43$}}  \\ \cline{3-12}
 & & SWME & $8.10$ & $4.00$ & $9.53$ & $1.80$ & {\setlength{\fboxsep}{1.5pt}\colorbox{2nd}{$25.06$}} & $1.57$ & $3.28$ & {\setlength{\fboxsep}{1.5pt}\colorbox{3rd}{$13.19$}} & {\setlength{\fboxsep}{1.5pt}\colorbox{1st}{$31.61$}}  \\ \cline{2-12}
 & \multirow{2}{*}{Excl.} & LAT.AVG. & $7.94$ & $3.92$ & {\setlength{\fboxsep}{1.5pt}\colorbox{3rd}{$9.34$}} & $1.53$ & $1.28$ & $1.68$ & $2.63$ & {\setlength{\fboxsep}{1.5pt}\colorbox{2nd}{$11.17$}} & {\setlength{\fboxsep}{1.5pt}\colorbox{1st}{$58.99$}}  \\ \cline{3-12}
 & & SWME & $6.61$ & $3.26$ & $7.78$ & $1.27$ & {\setlength{\fboxsep}{1.5pt}\colorbox{2nd}{$17.67$}} & $1.56$ & $2.19$ & {\setlength{\fboxsep}{1.5pt}\colorbox{3rd}{$9.29$}} & {\setlength{\fboxsep}{1.5pt}\colorbox{1st}{$49.10$}}  \\ \hline\hline
%\multirow{4}{*}{UTfit} & \multirow{2}{*}{incl.} & LAT.AVG. & $10.59$ & $5.23$ & {\setlength{\fboxsep}{1.5pt}\colorbox{3rd}{$12.47$}} & $1.70$ & $2.36$ & $1.98$ & $1.77$ & $2.87$ & {\setlength{\fboxsep}{1.5pt}\colorbox{2nd}{$16.42$}} & {\setlength{\fboxsep}{1.5pt}\colorbox{1st}{$41.49$}}  \\ \cline{3-13}
% & & SWME & $8.08$ & $3.99$ & $9.51$ & $1.28$ & $1.80$ & {\setlength{\fboxsep}{1.5pt}\colorbox{2nd}{$25.08$}} & $1.50$ & $2.19$ & {\setlength{\fboxsep}{1.5pt}\colorbox{3rd}{$12.53$}} & {\setlength{\fboxsep}{1.5pt}\colorbox{1st}{$31.65$}}  \\ \cline{2-13}
% & \multirow{2}{*}{excl.} & LAT.AVG. & $7.91$ & $3.91$ & {\setlength{\fboxsep}{1.5pt}\colorbox{3rd}{$9.32$}} & $1.07$ & $1.53$ & $1.28$ & $1.61$ & $1.76$ & {\setlength{\fboxsep}{1.5pt}\colorbox{2nd}{$10.60$}} & {\setlength{\fboxsep}{1.5pt}\colorbox{1st}{$59.05$}}  \\ \cline{3-13}
% & & SWME & $6.59$ & $3.25$ & $7.76$ & $0.88$ & $1.27$ & {\setlength{\fboxsep}{1.5pt}\colorbox{2nd}{$17.68$}} & $1.49$ & $1.46$ & {\setlength{\fboxsep}{1.5pt}\colorbox{3rd}{$8.83$}} & {\setlength{\fboxsep}{1.5pt}\colorbox{1st}{$49.16$}}  \\ \hline\hline
\end{tabular}
}
\caption{Error Fractions $\sigma_{i}^{2}/\sum\sigma_{j}^{2}$. The UTfit Wolfenstein parameters(W.P) show the same tendency.}
\label{table:error fractions}
\end{table}
%

%-------------
% In order to observe a $5\sigma$ difference between
% $\epsilon_{K}^{\text{Exp}}$ and $\epsilon_{K}^{\text{SM}}$, we need to
% reduce the error of $B_K$ from the current 5.45\% down to 1.81\%, and
% simutaneously to reduce the error of $V_{cb}$ from 2.53\% to 0.63\% if
% the average values do not change as they are.
%
% To estimate this target error we use the input set consists of
% CKMfitter's parameters, exclusive $V_{cb}$, SWME $B_{K}$
% and change only the errors for the last two parameters.
%
% Reducing $V_{cb}$ will make $\bar{\eta}$ be determined more precisely
% from the global fit.
%
% Therefore, if the SM fails to explain the elementary particle
% interactions, we can reach the $5\sigma$ difference before the target
% errors are achieved.
%-------------

% EDIT

%\newpage
\section{Acknowledgements}
The research of W.~Lee is supported by the Creative Research
Initiatives Program (2012-0000241) of the NRF grant funded by the
Korean government (MEST).
W.~Lee would like to acknowledge the support from KISTI supercomputing
center through the strategic support program for the supercomputing
application research [No. KSC-2011-G2-06].
Computations were carried out in part on QCDOC computing facilities of
the USQCD Collaboration at Brookhaven National Lab, on GPU computing
facilities at Jefferson Lab, on the DAVID GPU clusters at Seoul
National University, and on the KISTI supercomputers. The USQCD
Collaboration are funded by the Office of Science of the
U.S. Department of Energy.


\begin{thebibliography}{99}

%\bibitem{Cabibbo1963:PhysRevLett.10.531} N.~Cabibbo,
%  \emph{Phys.Rev.Lett.} {\bf 10} (1963) 531.

%\bibitem{Kobayashi1973:ProgTheorPhys.49.652} M.~Kobayashi, and
%  T.~Maskawa, \emph{Prog.Theor.Phys.} {\bf 49} (1973) 652.

\bibitem{Nakamura2010:RevPtlPhys} K.~Nakamura \textit{et. al.}
  (Particle Data Group), \emph{J.Phys.G} {\bf G37} (2010) 075021.

\bibitem{swme-prl-1} T.~Bae \textit{et. al.},
  \emph{Phys.Rev.Lett.} {\bf 109} (2012) 041601 [{\tt
    hep-lat/1111.5698}].

\bibitem{rbc-prd-2011-1} Y.~Aoki \textit{et. al.},
  \emph{Phys.~Rev.} {\bf D84} (2011) 014503 [{\tt
    hep-lat/1012.4178}].

\bibitem{alv-prd-2010-1} C.~Aubin \textit{et. al.},
  \emph{Phys.~Rev.} {\bf D81} (2010) 014507 [{\tt
    hep-lat/0905.3947}].

\bibitem{fnal/milc-pos-2010-1} J.~Bailey \textit{et. al.},
  \pos{PoS(LATTICE2010)311} [{\tt hep-lat/1011.2166}].

\bibitem{Christ2011:hep-lat/1201.2065} N.~Christ, in \emph{Proceedings of Lattice2011} \pos{PoS(LATTICE2011)277} [{\tt hep-lat/1201.2065}].

\bibitem{Buras2010:PhysLettB.688} A.~J.~Buras, D.~Guadagnoli and
  G.~Isidori, \emph{Phys. Lett.} {\bf B688} (2010) 309 [{\tt
    hep-ph/1002.3612}].

\bibitem{Buras1998:hep-ph/9806471} A.~J.~Buras, \emph{Weak
    Hamiltonian, CP violation and rare decays} in \emph{Probing the
    Standard Model of Particle Interactions}, F. David and R. Gupta,
  eds., Elsevier Science B.V. (1998) [{\tt hep-ph/9806471}].

\bibitem{Brod2011:PhysRevLett.108.121801} J.~Brod and M.~Gorbahn,
  \emph{Phys.Rev.Lett} {\bf 108} (2012) 121801 [{\tt hep-ph/1108.2036}].

\bibitem{Buras2008:PhysRevD.78.033005} A.~J.~Buras, and D.~Guadagnoli,
  \emph{Phys.Rev.} {\bf D78} (2008) 033005 [{\tt hep-ph/0805.3887}].

\bibitem{Laiho2009:PhysRevD.81.034503} J.~Laiho, and E.~Lunghi, and
  R.~S.~Van~de~Water, \emph{Phys.Rev.} {\bf D81} (2010) 034503 [{\tt
    hep-ph/0910.2928}] {\tt http://latticeaverages.org/}.

\bibitem{Laiho2011:hep-ex/1107.3934} J.~Laiho, and B.~D.~Pecjak, and
  C.~Schwanda, in \emph{Proceedings of CKM2010} (2011) [{\tt
    hep-ex/1107.3934}].

\bibitem{Blum2011:PhysRevLett.108.141601} T.~Blum \textit{et. al.},
  \emph{Phys.Rev.Lett.} {\bf 108} (2012) 141601 [{\tt
    hep-lat/1111.1699}].

\end{thebibliography}
\end{document}